\documentclass[runningheads]{llncs}
\usepackage{graphicx}
\usepackage{diagbox}
\usepackage{makecell}
\usepackage{multirow}
\usepackage{cite}
\usepackage{booktabs}
\usepackage{caption}
\usepackage{subcaption} 
\usepackage{amsmath,amssymb,amsfonts}
\usepackage{xcolor}
\usepackage{floatrow}
\newfloatcommand{capbtabbox}{table}[][\FBwidth]
\newcommand{\para}[1]{\vspace{.05in}\noindent\textbf{#1}}

\usepackage[colorlinks,linkcolor=blue]{hyperref}
\usepackage{marvosym}
\usepackage{makecell}
\usepackage[colorlinks,linkcolor=blue]{hyperref}
\authorrunning{Xing et al.}
\titlerunning{NestedFormer}

\begin{document}
\title{NestedFormer: Nested Modality-Aware Transformer for Brain Tumor Segmentation}

\author{Zhaohu Xing\inst{1} \and
Lequan Yu\inst{2}\and
Liang Wan\inst{1}\textsuperscript{(\Letter)}\and
Tong Han\inst{3} \and
Lei Zhu\inst{4,5}
}
\institute{Medical College of Tianjin University, 
Tianjin, China \\
\email{xingzhaohu@tju.edu.cn}, \email{lwan@tju.edu.cn} \and
The University of Hong Kong, Hong Kong, China \and
Brain Medical Center of Tianjin University, Huanhu Hospital, Tianjin, China \and
The Hong Kong University of Science and Technology (Guangzhou), Guangzhou, China \and
The Hong Kong University of Science and Technology, Hong Kong, China \
}

\titlerunning{Nested Modality-Aware Transformer}

\maketitle              

\begin{abstract}
Multi-modal MR imaging is routinely used in clinical practice to diagnose and investigate brain tumors by providing rich complementary information. 
Previous multi-modal MRI segmentation methods usually perform modal fusion by concatenating multi-modal MRIs at an early/middle stage of the network, which hardly explores non-linear dependencies between modalities.  
In this work, we propose a novel Nested Modality-Aware Transformer (NestedFormer) to explicitly explore the intra-modality and inter-modality relationships of multi-modal MRIs for brain tumor segmentation.
Built on the transformer-based multi-encoder and single-decoder structure, we perform nested multi-modal fusion for high-level representations of different modalities and apply modality-sensitive gating (MSG) at lower scales for more effective skip connections. 
Specifically, the multi-modal fusion is conducted in our proposed Nested Modality-aware Feature Aggregation (NMaFA) module, which enhances long-term dependencies within individual modalities via a tri-orientated spatial-attention transformer, and further complements key contextual information among modalities via a cross-modality attention transformer. 
Extensive experiments on BraTS2020 benchmark and a private meningiomas segmentation (MeniSeg) dataset show that the NestedFormer clearly outperforms the state-of-the-arts.
The code is available at \href{https://github.com/920232796/NestedFormer}{https://github.com/920232796/NestedFormer}.

\end{abstract}

\section{Introduction}

\begin{figure}[t]
\centering
\includegraphics[width=\textwidth]{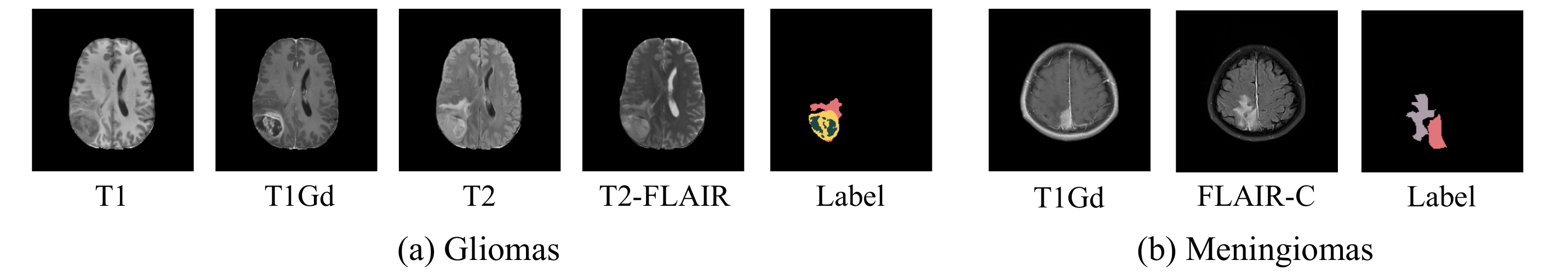}
\caption{Multi-modal MRIs for (a) Gliomas; and (b) Meningiomas.} 
\label{fig:MMdata}
\vspace{-2mm}
\end{figure}

Brain tumor is one of the most common cancers in the world~\cite{Bray2018Cancers}, in which gliomas are the most common malignant brain tumors with different levels of aggressiveness and meningiomas are the most prevalent primary intracranial tumors in adults~\cite{Ostrom2020BrainTumor}. 
Multi-modal magnetic resonance imaging (MRI) is routinely used in the clinic by providing rich complementary information for analyzing brain tumors.
Specifically, for gliomas, the commonly used MRI sequences are T1-weighted (T1), post-contrast  T1-weighted (T1Gd), T2-weighted (T2) and T2 Fluid Attenuation Inversion Recovery (T2-FLAIR) images; see Fig.~\ref{fig:MMdata}(a), each with varying roles in distinguishing tumor, peritumoral edema and tumor core ~\cite{menze2014multimodal, bakas2017advancing,bakas2018identifying}.
For meningiomas, they have different characteristic appearances on T1Gd~\cite{li2019presurgical} and contrast-enhanced T2-FLAIR (shorted for FLAIR-C) MRI images; see Fig.~\ref{fig:MMdata}(b).
Thus, automatic segmentation of brain tumor structures from multi-modal MRIs is important for clinical diagnosis and treatment planning.

In recent years, convolutional neural networks (CNNs) have achieved promising successes in brain tumor segmentation. 
The main stream models are built upon the encoder-decoder architecture~\cite{ronneberger2015u} with skip connections, including S3D-UNet~\cite{chen2018s3d}, SegResNet~\cite{myronenko20183d}, HPU-Net~\cite{kong2018hybrid}, etc. 
Recent works~\cite{wang2021transbts, zhang2021DualTrans, hatamizadeh2022unetr} also explore transformer~\cite{vaswani2017attention} to model long-range dependencies within images.
For instance, TransBTS~\cite{wang2021transbts} utilizes 3D-CNN to extract local spatial features, and applies transformer to model global dependencies on high-level features. 
UNETR~\cite{hatamizadeh2022unetr} uses the ViT transformer as the encoder to learn contextual information, which is merged with the CNN-based decoder via skip connections at multiple resolutions.
However, the transformer in these methods is used to enhance the encode path without specific design for multi-modal fusion.

To utilize the multi-modal information, most of existing methods adopt an early-fusion strategy, in which multi-modal images are concatenated as the network input.
However, this strategy can hardly explore non-linear relationships between different modalities.
To alleviate this problem, recent works follow a layer-fusion strategy~\cite{Dolz2019HyperDense, zhou20213d, zhang2021modality}, where the modality-specific features extracted by different encoders are fused in the middle layers of the network and share the same decoder. 
In HyeprDense-Net~\cite{Dolz2019HyperDense}, each modality has a separated stream and dense connections are introduced between layers within the same stream and also across different streams. 
MAML~\cite{zhang2021modality} embeds multi-modal images by different modality-specific FCNs and then applies a modality-aware module to regress attention maps in order to fuse the modality-specific features. 
Nevertheless, these multi-modal fusion methods do not build the long-range spatial dependencies within and cross modalities, so that they cannot fully utilize the complementary information of different modalities.

In this paper, we propose a novel nested modality-aware transformer, called NestedFormer, for effective and robust multi-modal brain tumor segmentation.
We first design an effective Global Poolformer to extract discriminative volumetric spatial features, with more emphasis on global dependencies, from different MRI modalities. 
To better extract the complementary features and enable any number of modalities for fusion, we propose a novel Nested Modality-aware Feature Aggregation (NMaFA) module.
It explicitly considers both single-modality spatial coherence and cross-modality coherence, and leverages nested transformers to establish the intra- and inter-modality long-range dependencies, resulting in more effective feature representation. 
Moreover, we design a computationally efficient Tri-orientated Spatial Attention (TSA) paradigm to accelerate the 3D-spatial-coherence calculation.
To improve feature reuse effect in the decoding, a novel modality-sensitive gating (MSG) module is developed to dynamically filter modality-aware low-resolution features for effective skip connections. 
Extensive experiments on BraTS2020 benchmark and a privately collected meningiomas segmentation dataset (MeniSeg) show that our model clearly ourperforms the state-of-the-art methods.
\begin{figure}[t]
\centering
\includegraphics[width=\textwidth]{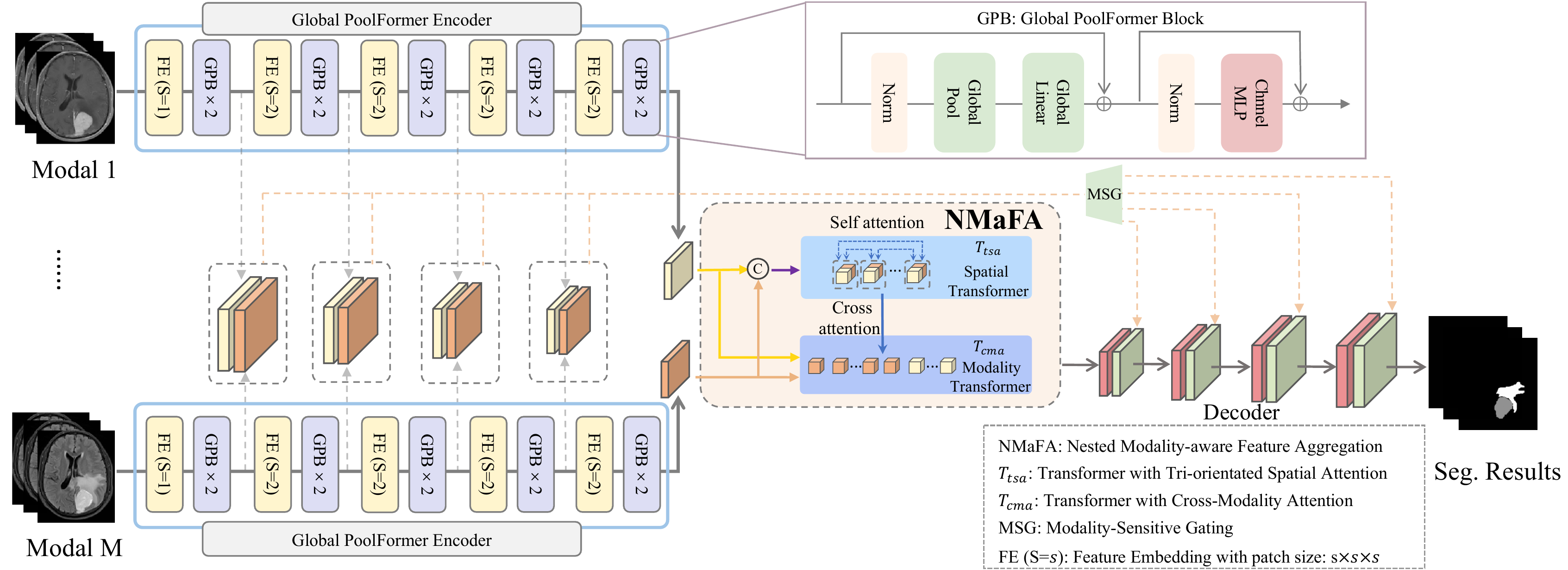}
\caption{An overview of the proposed NestedFormer. We design a Nested Modality-aware Feature Aggregation (NMaFA) module to model both the intra- and inter-modality features for multi-modal fusion.} 
\label{fig:framework}
\vspace{-1mm}
\end{figure}

\vspace{-1mm}
\section{Method}
\vspace{-1mm}

Fig.~\ref{fig:framework} illustrates the overview of the proposed NestedFormer, which consists of three components: 1) multiple encoders to obtain multi-scale representations of different modalities, 2) a NMaFA fusion module to explore correlated features within and between multi-modal high-level embeddings, and 3) a gating strategy to selectively transfer modality-sensitive low-resolution features to the decoder. 

\vspace{-1mm}
\subsection{Global Poolformer Encoder}
\vspace{-1mm}
Recent works show that transformer is more conducive to modeling global information than CNNs.
To better extract local context information for each modality, we extend the Poolformer~\cite{yu2021metaformer} as the modality-specific encoder. 
As discussed in~\cite{yu2021metaformer}, replacing the computation-intensive attention module in Transformer with average pooling can achieve superior performance than recent transformer and MLP-like models.
Therefore, to enhance global information, we design Global Poolformer Block (GPB), which leverages global pooling instead of average pooling in Poolformer, followed by a fully connected layer.
As shown in Fig.~\ref{fig:framework}, given the input feature embedding $X$, a GPB block consists of a learnable global pooling (GP) and a MLP sub-block. The output $Z$ is computed as,
\vspace{-1mm}
\begin{equation}\label{Equ:gpoolformer}
    \begin{aligned}
    Y  =  GP(LN(X))W_g + X,\\
    Z  =  MLP(LN(Y)) + Y,
    \end{aligned}
\vspace{-1mm}
\end{equation}
where $LN(*)$ denotes the layer normalization and $W_g$ is the learnable parameter in the FC layer. 
Our Global Poolformer encoder contains five groups of one feature embedding (FE) layer and two GPB blocks. 
Each FE layer is a 3D-convolution, while the first FE layer has a convolution patch size of $1\times1\times1$ and the rest layers have a patch size of $2\times2\times2$ and a stride of 2.
The encoders gradually encode each modality image into high-level feature $F_{L, i} \in \mathbb{R}^{d \times w \times h \times C}, i\in [1,M]$, where $(d, w, h)=(\frac{D}{16}, \frac{W}{16}, \frac{H}{16})$ are $1/16$ of input spatial resolutions $H$, $W$ and depth dimension $D$; $M$ is the number of modal images, the channel dimension $C$ and the layer number $L$ are set as $C=128$, $L=5$.

\vspace{-2mm}
\subsection{Nested Modality-Aware Feature Aggregation}
\begin{figure}[t]
\centering
\includegraphics[width=\textwidth]{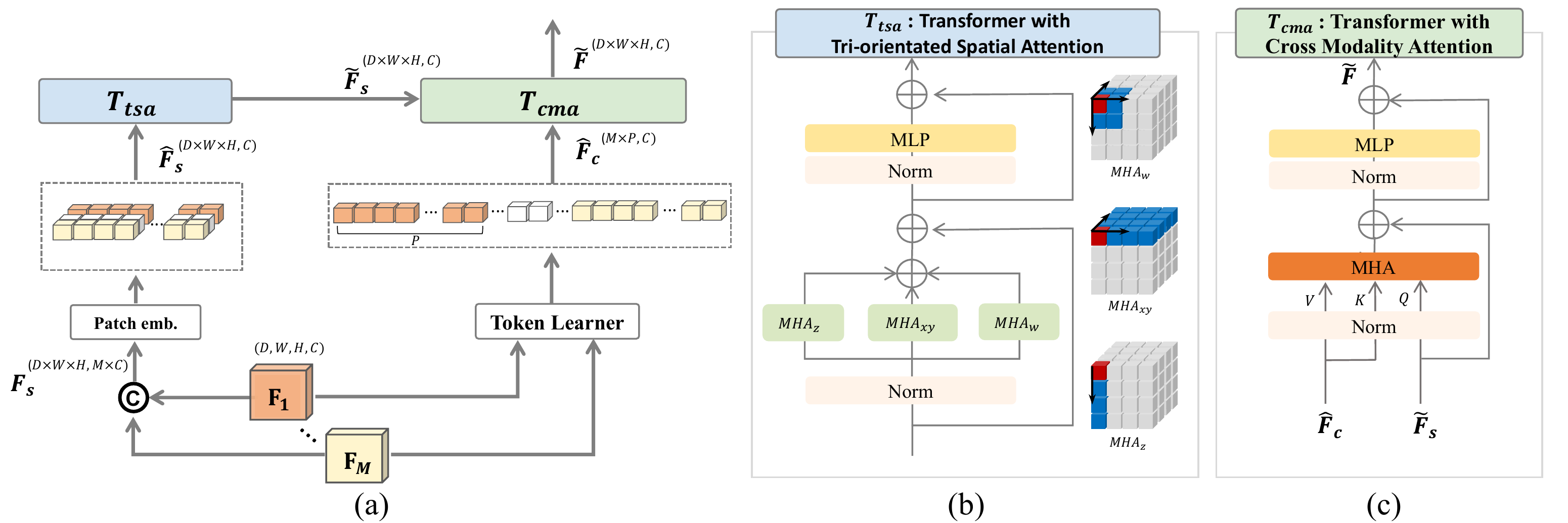}
\caption{NMaFA: Nested Modality-aware Feature Aggregation. (a) The overall architecture. (b) The transformer with tri-orientated spatial attention $T_{tsa}$. (c) The transformer with cross-modality attention $T_{cma}$.}
\label{fig:LDaFA}
\vspace{-1mm}
\end{figure}

\vspace{-1mm}
Given high-level features  $F_{L, i}, i\in [1,M]$, 
NMaFA leverages a spatial-attention based transformer $T_{tsa}$ and a cross-modality attention based transformer $T_{cma}$ in a nested manner; see Fig.~\ref{fig:LDaFA}. 
First, transformer $T_{tsa}$ utilizes the self-attention scheme to compute the long-range correlation between different patches in the space within each modality.
Specifically, $F_{L, i}$ is \textit{concatenated in the channel dimension} to obtain high-level embedding $F_s \in \mathbb{R}^{d \times w \times h \times MC}$. 
In this work, each location of $F_s$ is considered as one ``patch''. Then a patch embedding layer maps $F_s$ to a token sequence $\hat{F}_s \in \mathbb{R}^{dwh\times C}$.
$T_{tsa}$ takes $\hat{F}_s$ and the position encoding~\cite{vaswani2017attention} as the input, and outputs spatially-enhanced feature $\tilde{F}_s \in \mathbb{R}^{dwh\times C}$. 

Second, transformer $T_{cma}$ utilizes the cross-attention scheme to further compute the global relation among different modalities to achieve inter-modality fusion.
To this end, $F_{L,i}$ is \textit{concatenated in the spatial dimension} to obtain the flatten sequence $\hat{F}_c \in \mathbb{R}^{MP \times C}$. 
Here, $P (P = 32)$ denotes the number of dominant tokens learnt via the Token Learner strategy~\cite{ryoo2021tokenlearner}, which helps to reduce the computational scope especially when the number of tokens increases greatly along with more modalities. 
After that, both $\tilde{F}_s$ and $\hat{F}_c$ are fed into $T_{cma}$ to get the modality-enhanced feature embedding $\tilde{F}$.

Also note that our two modules are different from traditional channel-spatial attention networks, which reweigh feature maps channel-wise and spatial-wise. Our NMaFA relies on transformer mechanism and the two transformers are fused in a nested form, rather than serial~\cite{khanh2020enhancing} or parallel~\cite{mou2019cs} fusion.

\para{Transformer with Tri-orientated Spatial Attention.}
To improve the computational efficiency of spatial attention for volumetric embeddings, inspired by \cite{ho2019axial,liu2021swin}, we leverage axial-wise attention $MHA_{z}$, plane-wise attention $MHA_{xy}$, and window-wise attention $MHA_{w}$; see Fig.~\ref{fig:LDaFA}(b).
Concretely, $MHA_{z}$ models the long-range relationship among feature tokens along the vertical direction; 
$MHA_{xy}$ models the long-range relationship within each slice;
$MHA_{w}$ uses sliding windows to model the relationship across local 3D-windows. 
We employ axial and planar learnable absolute position encodings~\cite{vaswani2017attention} for $MHA_{z}$ and $MHA_{xy}$, respectively, and use relative position encoding for window-wise attention $MHA_{w}$~\cite{liu2021swin}. 
The resultant attention is computed as follows,
\vspace{-1mm}
\begin{equation}\label{Equ:multi_spatial_attention_2}
    \begin{aligned}
    MHA_{tsa}(z)=MHA_{z}(z) + MHA_{xy}(z) + MHA_{w}(z),
    \end{aligned}
\vspace{-1mm}
\end{equation}
where $z \in \mathbb{R}^{N \times C}$ denotes the embedding tokens with sequence length $N$ and embedding dimension $C$ after layer normalization, $N=d \times w \times h$.
By this way, the model not only enhances feature extraction of local important regions, but also calculates global feature dependencies with less computation.

\para{Transformer with Cross-Modality Attention.}
By concatenating features \textit{in the channel dimension}, $T_{tsa}$ mainly enhances the dependencies within each modality and yields $\tilde{F}_{s}$, although the inter-modality integration also takes place via patch embedding. 
To explicitly explore relationship among modalities, we concatenate the feature tokens of different modalities \textit{along the spatial dimension}, yielding $\hat{F}_c \in \mathbb{R}^{MP \times C}$; and then use a cross-attention transformer $T_{cma}$ to enhance the modality dependency information into $\tilde{F}_{s}$; see Fig.~\ref{fig:LDaFA}(c). 
The input triplet of (Query, Key, Value) to the cross-attention is computed as
\vspace{-1mm}
\begin{equation}\label{Equ:QKV}
    \begin{aligned}
    Q = \tilde{F}_{s}W_q, \; 
    K = \hat{F}_c W_k, \;
    V = \hat{F}_c W_v, 
    \end{aligned}
\vspace{-1mm}
\end{equation}
where $W_q, W_k, W_v \in \mathbb{R}^{C\times d}$ are the weight matrices, $d=128$ is the dimension of $Q, K, V$. The cross attention $CA$ is then formulated as
\vspace{-1mm}
\begin{equation}\label{Equ:TCA}
    \begin{aligned}
    CA(\tilde{F}_{s},\hat{F}_c)=\hat{F}_s + SoftMax(\frac{QK^{T}}{\sqrt{d}})V.
    \end{aligned}
\vspace{-1mm}
\end{equation}
The resultant token sequence $\tilde{F}$ from $T_{cma}$ fuses and enhances the input features with increasing receptive ﬁelds and the cross-modal global relevance.

\vspace{-1mm}
\subsection{Modality-Sensitive Gating}
\vspace{-1mm}
In feature decoding, we first fold the tokens $\tilde{F}$ back to a high-level 4D feature map $R_L \in \mathbb{R}^{d \times w \times h \times C}$.
$R_L$ is progressively processed in a regular bottom-up style with a 3D convolution and $2 \times$ upsampling operation to recover a full resolution feature map $R_1 \in \mathbb{R}^{D \times W \times H \times N_c}$ for segmentation, where $N_c$ is the number of segments. 
Note that the encoder features are multi-modal. 
Hence, we design a modality-sensitive gating strategy in skip connection, to filter the encoder features $\{F_{l, i}, l\in[1,4],i\in[1,M]\}$ according to the modality importance.
To be specific, for the $l$-th layer, an modality importance map $I_l \in \mathbb{R}^{ \frac{D}{2^{l-1}} \times \frac{W}{2^{l-1}} \times \frac{H}{2^{l-1}} \times M}$ is learnt from $\tilde{F}$ that is the output of NMaFA, as follows,
\vspace{-1mm}
\begin{equation}\label{equ:select}
    \begin{aligned}
    I_{l} = \sigma(U^{2\times}_{L-l}(FC(\tilde{F}))),
    \end{aligned}
\vspace{-1mm}
\end{equation}
where $FC(*)$ is a $1\times1\times1$ full connection layer, $U_{l}^{2\times}$ denotes upsampling $l$ times, and $\sigma(*)$ is the sigmoid function.
Denote $\odot$ as element-wise multiplication. Then the filtered encoder feature $\mathcal{F}_l$ is formulated as
\vspace{-1mm}
\begin{equation}\label{equ:filter}
    \begin{aligned}
    \mathcal{F}_{l} = \sum_i(I_{l,i} \odot F_{l, i}).
    \end{aligned}
\vspace{-1mm}
\end{equation}

\vspace{-4mm}
\section{Experiment}
\vspace{-1mm}
\subsection{Implementation Details}
\vspace{-1mm}
Our NestedFormer was implemented in PyTorch1.7.0 on a NVIDIA GTX 3090 GPU. 
The parameters were initialized via Xavier~\cite{glorot2010understanding}.
The loss function was a combination of soft dice loss and cross-entropy loss and we adopted the AdamW optimizer~\cite{loshchilov2017decoupled} with a weight decay of $10^{-5}$. 
The learning rate was empirically set as $10^{-4}$.
We adopted two $T_{tsa}$ sequentially and just one $T_{cma}$. In $MHA_w$, the window-size was set as (2, 2, 2) for BraTS2020 and (2, 4, 4) for MeniSeg.

\vspace{-2mm}
\subsection{Datasets and Evaluation Metrics}
\vspace{-1mm}
For evaluation, we use a public brain tumor segmentation dataset \textbf{BraTS2020}~\cite{menze2014multimodal} and a private 3D meningioma segmentation dataset (\textbf{MeniSeg}) collected from Brain Medical Center of Tianjin University, Tianjin Huanhu hospital.
Dice score and 95\% Hausdorff Distance (HD95) are adopted for quantitative comparison.

\para{BraTS2020 Dataset.} \ 
The BraTS2020 training dataset contains 369 aligned four-modal MRI data (i.e., T1, T1Gd, T2, T2-FLAIR), with expert segmentation masks (i.e., GD-enhancing tumor, peritumoral edema, and tumor core). 
Each modality has a $155\times 240 \times 240$ volume and is already resampled and co-registered. 
The segmentation task aims to segment the whole tumor (WT), enhancing tumor (ET), and tumor core (TC) regions. 
Following the recent work~\cite{larrazabal2021orthogonal}, we randomly divide the dataset into training (315), validation (17) and test (37). 

\para{Meningioma Dataset.} \ 
The MeniSeg dataset contains $110$ annotated two-modal MRIs (i.e., T1Gd and FLAIR-C) from the meningiomas patients, who had undergone tumor resection between March 2016 and March 2021. 
MRI scans were performed with four 3.0T MRI scanners (Skyra, Trio, Avanto, Prisma from Siemens).
Two radiologists annotated meningioma tumor and edema masks on T1Gd and FLAIR-C MRIs, and the third high-experienced radiologist made examination.
Each modality data has a volume of $32\times256\times256$, and is aligned into the same space and sampled to volume sizes of [32, 192, 192] for training. 
Two-fold cross-validation is conducted for all the compared methods.

\begin{table*}[t]
    \centering
    \caption{Quantitative comparison on BraTS 2020 dataset.}
    \vspace{-3mm}
    \label{tab:two_dataset_result}
    \renewcommand\arraystretch{1.3}
    \setlength\tabcolsep{5pt}
    \resizebox{\textwidth}{!}{
    \begin{tabular}{c |c c c | c c c c c c  c c  c| c c c}
    \Xhline{1pt}
    \multirow{2}{*}{Methods} & \multirow{2}{*}{\makecell{Param\\(M)}} & \multirow{2}{*}{\makecell{FLOPs\\(G)}} &  & \multicolumn{2}{c}{WT} &  & \multicolumn{2}{c}{TC} & &  \multicolumn{2}{c}{ET} &  & & \multicolumn{2}{c}{Ave} \\
    \cline{5-6} \cline{8-9} \cline{11-12} \cline{15-16} 
     & & & & Dice$\uparrow$ & HD95$\downarrow$ & & Dice$\uparrow$ & HD95$\downarrow$ & & Dice$\uparrow$ & HD95$\downarrow$ & & & Dice$\uparrow$ & HD95$\downarrow$  \\
    \Xhline{1pt}
    3D-UNet~\cite{cciccek20163d} & 5.75 & 1449.59 & & 0.882 & 5.113 &  & 0.830 & 6.604 &  & 0.782 & 6.715 &  & & 0.831 & 6.144  \\
    \hline
    SegResNet~\cite{myronenko20183d} & 18.79 & 185.23  & & 0.903 & 4.578 &  & 0.845 & 5.667 &  & 0.796 & 7.064 &  & & 0.848 & 5.763\\
    \hline
    MAML~\cite{zhang2021modality} & 5.76 & 577.65  & &  0.914 & 4.804 &  & 0.854 & 5.594 &  & 0.796 & \textbf{5.221} &  & & 0.855 & 5.206 \\
    \hline
    nnUNet~\cite{isensee2021nnu} & 5.75 & 1449.59  & &  0.907 & 6.94 &  & 0.848 & 5.069 &  & \textbf{0.814} & 5.851 &  & & 0.856 & 5.953 \\
    \Xhline{1pt}
    SwinUNet(2D)~\cite{cao2021swin} & 27.17 & 357.49  & & 0.872 & 6.752 &  & 0.809 & 8.071 &  & 0.744 & 10.644 &  & & 0.808 & 8.489 \\
    \hline
    TransBTS~\cite{wang2021transbts} & 32.99 & 333  & & 0.910 & \textbf{4.141} &  & 0.855 & 5.894 &  & 0.791 & 5.463 &  & & 0.852 & 5.166  \\
    \hline
    UNETR~\cite{hatamizadeh2022unetr} & 92.58 & 41.19  & & 0.899 & 4.314 &  & 0.842 & 5.843 &  & 0.788 & 5.598 &  & & 0.843 & 5.251  \\
    \Xhline{1pt}
    NestedFormer & 10.48 & 71.77 & & \textbf{0.920} & 4.567 &  & \textbf{0.864} & \textbf{5.316} &  & 0.800 & 5.269 &  & &  \textbf{0.861} & \textbf{5.051}\\
    \Xhline{1pt}
    \end{tabular}
    }
\vspace{-5mm}
\end{table*}

\subsection{Comparison with SOTA Methods}
We compare our network against 
seven SOTA segmentation methods, including three CNN-based methods (3D-UNet~\cite{cciccek20163d}, SegResNet~\cite{myronenko20183d}, MAML~\cite{zhang2021modality}, nnUNet~\cite{isensee2021nnu}),
and three transformer-based methods 
(SwinUNet(2D)~\cite{cao2021swin},
TransBTS~\cite{wang2021transbts},
and UNETR~\cite{hatamizadeh2022unetr}). 
For a fair comparison, we utilize the public implementations of compared methods to re-train their networks for generating their best segmentation results. 
Considering the computation power, all the methods are trained for at most 300 epochs on BraTS2020 and 200 epochs on MeniSeg.

\para{BraTS2020.} Table~\ref{tab:two_dataset_result} reports the Dice and HD95 scores on three regions (WT, TC, and ET) as well as the averaged scores of all the methods on BraTS2020.
Apparently, our NestedFormer achieves the largest Dice score on WT, the largest Dice score on TC, the smallest HD95 scores on TC, 
and our method also ranks second in Dice score on ET, and second in HD95 score on WT and ET.
More importantly, our method has the best quantitative performance with averaging Dice and HD95 scores to be 0.861 and 5.051. 
It is noted that HD95 is for the distance difference between two sets of points, which is more sensitive than Dice~\cite{wang2021transbts}. Hence, Dice is often used as the main metric and HD95 as the reference. 
We also experimented with two-fold cross-validation for UNETR, TransBTS and our method, while our method outperforms the two methods in WT and TC, and is quite close to the best result in ET.
As for model complexity, our model has 10.48M parameters and 71.77G FLOPs which is a moderate size model. 

\para{MeniSeg.}
In Table~\ref{tab:MeniSeg_result}, we list Dice and HD95 scores of our network and compared methods on tumor and edema regions on the MeniSeg dataset, as well as the average metrics.
Among all the compared methods, MAML has the largest Dice score of 0.819 at the  tumor segmentation, while UNETR has the largest Dice score of 0.693 at the edema segmentation, and average Dice score of 0.755.
In comparison, our method has a 1.5\% Dice improvement in meningioma tumor, 0.2\% Dice improvement in edema, and 1.0\% average Dice improvement.
Regarding HD95, our method achieves the 4th smallest score of 2.647 on the tumor segmentation, and the smallest score of 6.173 on the edema segmentation. 

\begin{table*}[tp]
\begin{floatrow}
\resizebox{0.5\textwidth}{!}{
\centering
\renewcommand\arraystretch{1.6}
\ttabbox{\caption{Quantitative comparison on MeniSeg dataset.}}{%
\vspace{-2mm}
\label{tab:MeniSeg_result}
\begin{tabular}{c | c c c c c c | c c c }
    \toprule
    \multirow{2}{*}{Methods} & \multicolumn{2}{c}{Tumor} &  & \multicolumn{2}{c}{Edema} &  & & \multicolumn{2}{c}{Ave}\\
    \cline{2-3} \cline{5-6} \cline{9-10}
     & Dice & HD95 &  & Dice & HD95 & & & Dice & HD95 \\
    \Xhline{1pt}
    3D-UNet~\cite{cciccek20163d} & 0.799 & 5.099 &  & 0.676 & 9.655 &  & & 0.737 & 7.377 \\
    \hline
    SegResNet~\cite{myronenko20183d} & 0.813 & 2.970 &  & 0.665 & 10.438 &  & & 0.739 & 6.704\\
    \hline
    MAML~\cite{zhang2021modality} &  0.819 & 2.112 &  & 0.682 & 9.158 &  & & 0.750 & 5.635\\
    \Xhline{1pt}
    SwinUNet(2D)~\cite{cao2021swin} & 0.807 & 1.817 &  & 0.679 & 7.986 &  & & 0.743 & 4.901\\
    \hline
    TransBTS~\cite{wang2021transbts} & 0.809 & \textbf{1.742} &  & 0.679 & 6.388 &  & & 0.744 & \textbf{4.065}\\
    \hline
    UNETR~\cite{hatamizadeh2022unetr} & 0.818 & 3.279 &  & 0.693 & 7.837 &  & & 0.755 & 5.813\\
    \Xhline{1pt}
 
    NestedFormer & \textbf{0.834} & 2.647 &  & \textbf{0.695} & \textbf{6.173} &  & & \textbf{0.765} & 4.410\\
    \Xhline{1pt}
    \end{tabular}
    }
}

\resizebox{0.5\textwidth}{!}{
\centering
\renewcommand\arraystretch{2.0}
\begin{floatrow}
\ttabbox{\caption{Ablation study for different modules on MeniSeg.}
\vspace{-2mm}
\label{tab:ablation_result}}{%
\begin{tabular}{c| c c c c c c c  c |c c c c c c}
    \Xhline{1pt}
     \multirow{2}{*}{Methods} & \multicolumn{3}{c}{Encoder} &  & \multicolumn{3}{c}{Fusion} &   & & \multicolumn{3}{c}{Dice}\\
    \cline{2-4} \cline{6-8} \cline{11-13}
     & CNN & PB & GPB &  &  $T_{tsa}$ & $T\_{cma}$ & MSG & & & Tmuor & Edema & Ave\\
    \Xhline{1pt}
   
    baseline 1 & {\checkmark} &  &  &  &  &  &  &  &  & 0.805 & 0.675 & 0.74 \\
    \hline
    baseline 2 &  &  & {\checkmark} &  &  &  &  &  & & 0.810 & 0.679 & 0.75 \\
    \hline
    baseline 3 &  &  & {\checkmark} &  & {\checkmark} &  &  &  & & 0.816 & 0.688 & 0.752 \\
    \hline
    baseline 4 &  &  & {\checkmark} &  & {\checkmark} & {\checkmark} &  &  & & 0.825 & \textbf{0.699} & 0.762 \\
    \hline
    baseline 5 &  & {\checkmark} &  &  & {\checkmark}
    & {\checkmark} & {\checkmark} &  & & 0.823 & 0.697 & 0.76 \\
    \hline
    NestedFormer &  &  & \checkmark &  & \checkmark
    & \checkmark & \checkmark &  & & \textbf{0.834} & 0.695 & \textbf{0.765}\\
    \Xhline{1pt}
    \end{tabular}
    }
\end{floatrow}
}
\end{floatrow}
\end{table*}

\begin{figure}[t]
\centering
\includegraphics[width=\textwidth]{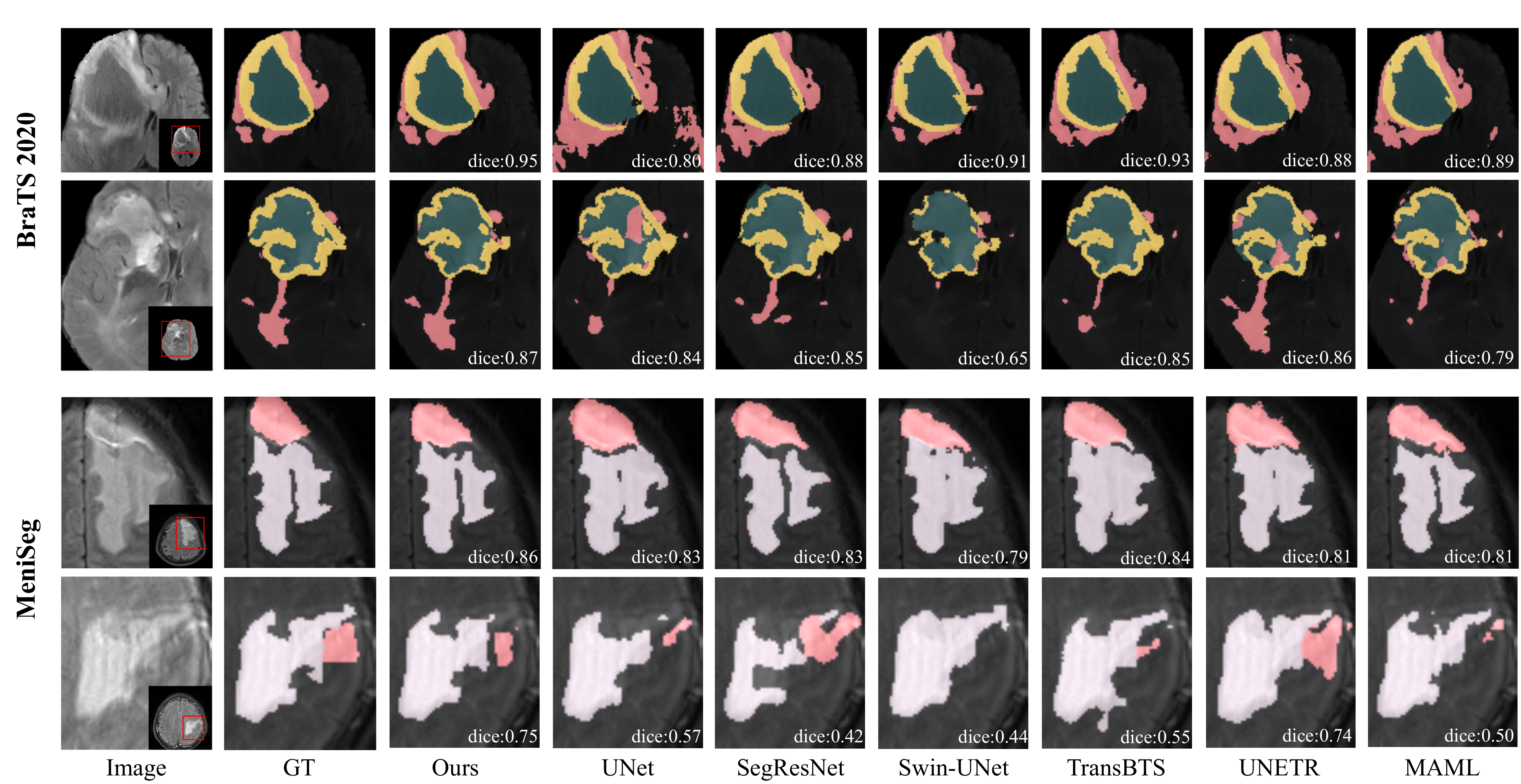}
\caption{The visual comparison results on BraTS2020 and MeniSeg dataset.}
\label{fig:visual_image}
\end{figure}

\para{Visual Comparisons on BraTS2020 and MeniSeg.}
Fig.~\ref{fig:visual_image} visually compares the segmentation results predicted by our network and SOTA methods on BraTS2020 and MeniSeg.
From these visualization results, we can find that our method can more accurately segment brain tumor and peritumoral edema regions than all the compared methods.
The reason behind is that our method is able to better fuse multi-modal MRIs by explicitly exploring the intra-modality and the inter-modality relationships among multiple modalities. 

\vspace{-2mm}
\subsection{Ablation study}
\vspace{-1mm}
We conduct ablation studies on the MeniSeg dataset to evaluate the contributions of main modules in our method; see Table~\ref{tab:ablation_result}.
We not only compared the effects of three different encoder backbones based on CNN, PB, and GP, but also verified the effect of our proposed fusion modules.
Among them, $baseline1$ uses multiple U-Net encoders to extract features of different modal images, and performs feature fusion by concatenation. 
$baseline2$-$baseline4$ uses multiple GPB encoders to extract features, and conducts skip connection via simple convolution, w/o $T_{tsa}$ and $T_{cma}$ (see Fig.~\ref{fig:LDaFA}), respectively. 
$baseline5$ replaces GPB block with the original PoolFormer block (PB) in the encoder, using the proposed NMaFA module (including $T_{tsa}$ and $T_{cma}$) as well as MSG.  
It can be observed clearly that compared with $baseline2$, using the NMaFA module enhances the extraction of long-distance dependency information and effectively improves the segmentation results, while GPB outperforms PB by considering global information.
Moreover, the MSG module is added to increase the feature reuse capability of skip connections, which further improves the segmentation effect, achieving the best average segmentation Dice (0.765) on the MeniSeg dataset.

\vspace{-1mm}
\section{Conclusion}
\vspace{-1mm}
We propose a novel multi-modal segmentation framework, dubbed as NestedFormer.
This architecture extracts the features of $M$ modalities by using multiple Global Poolformer Encoders. Then, the high-level features are effectively fused by the NMaFA module, and the low-level features are selected by the modality-sensitive gate (MSG) module.
Through these proposed modules, the network effectively extracts and hierarchically fuses features from different modalities. The effectiveness of our proposed NestedFormer is validated on BraTS2020 and MeniSeg datasets.
Our framework are modality-agnostic and can be extended to other multimodal medical data. In the future work, we will explore more efficient feature fusion on low-levels to further improve the segmentation performance.

\para{Acknowledgments.}
This work was supported by the grant from Tianjin Natural Science Foundation (Grant No. 20JCYBJC00960) and HKU Seed
Fund for Basic Research (Project No. 202111159073).

\bibliographystyle{splncs04}
\bibliography{mybibliography}

\end{document}